\documentclass[sigconf,screen]{acmart}
\usepackage{enumitem}
\usepackage{booktabs}
\usepackage{multirow}
\usepackage[ruled,linesnumbered,vlined]{algorithm2e} 

\AtBeginDocument{%
  }

\copyrightyear{2026}
\acmYear{2026}
\setcopyright{cc}
\setcctype{by}
\acmConference[DAC '26]{63rd ACM/IEEE Design Automation Conference}{July 26--29, 2026}{Long Beach, CA, USA}
\acmBooktitle{63rd ACM/IEEE Design Automation Conference (DAC '26), July 26--29, 2026, Long Beach, CA, USA}
\acmDOI{10.1145/3770743.3803942}
\acmISBN{979-8-4007-2254-7/2026/07}

\begin{document}
\newcommand{\papername}[1]{CellE}
\newcommand{\ry}[1]{\textcolor{blue}{[RY]: {#1}}}
\newcommand{\todo}[1]{\textcolor{blue}{#1}}
\setlist{leftmargin=0.5cm}
\setlength{\textfloatsep}{2pt}
\setlength{\floatsep}{2pt}
\title{\papername{}: Automated Standard Cell Library Extension via Equality Saturation}

\author{Yi Ren\textsuperscript{1,2,$\dagger$}, 
    Yukun Wang\textsuperscript{3}, 
    Xiang Meng\textsuperscript{3}, 
    Guoyao Cheng\textsuperscript{4}, 
    Baokang Peng\textsuperscript{4} \\
    Lining Zhang\textsuperscript{4,5}, 
    Yibo Lin\textsuperscript{1,5,6},
    Runsheng Wang\textsuperscript{1,5,6},
    Guangyu Sun\textsuperscript{1,5,6,*}}
\affiliation{%
        \textsuperscript{1}School of Integrated Circuits, 
        \textsuperscript{2}School of Software and Microelectronics, Peking University, Beijing, China \\
        \textsuperscript{3}School of Electronics Engineering and Computer Science, Peking University, Beijing, China \\
        \textsuperscript{4}School of Electronic and Computer Engineering, Peking University, Shenzhen, China \\
        \textsuperscript{5}Institute of Electronic Design Automation, Peking University, Wuxi, China \\
        \textsuperscript{6}Beijing Advanced Innovation Center for Integrated Circuits, Beijing, China \\
 \country{}
}%
\thanks{This work is supported in part by New Generation Artificial Intelligence-National Science and Technology Major Project (2025ZD0122105), Beijing Natural Science Foundation (L243001), National Natural Science Foundation of China (62125401), National Key Research and Development Program of China (2021ZD0114702), Beijing Outstanding Young Scientist Program (JWZQ20240101004) and 111 Project (B18001).}
\thanks{\textsuperscript{$\dagger$}First author, e-mail: \href{mailto:yiren20@stu.pku.edu.cn}{\texttt{yiren20@stu.pku.edu.cn}}}
\thanks{\textsuperscript{*}Corresponding author, e-mail: \href{mailto:gsun@pku.edu.cn}{\texttt{gsun@pku.edu.cn}}}

\renewcommand{\authors}{Yi Ren, Yukun Wang, Xiang Meng, Guoyao Cheng, Baokang Peng, Lining Zhang, Yibo Lin, Runsheng Wang, and Guangyu Sun}
\renewcommand{\shortauthors}{Ren et al.}

\begin{abstract}

Automated standard cell library extension is crucial for maximizing Quality of Results (QoR) in modern VLSI design. We introduce \papername{}, a novel framework that leverages formal methods to achieve exhaustive discovery of functionally equivalent subcircuits. \papername{} applies equality saturation to the post-mapping netlist, generating an e-graph to cluster all functionally equivalent implementations. This canonical representation enables an efficient pattern mining algorithm to select the most area-optimal standard cells. Experimental results show a 15.41\% average area reduction (up to 23.64\% over prior work). Furthermore, characterization in a commercial flow demonstrates an 8.00\% average delay reduction, confirming \papername{}'s superior QoR optimization capabilities.

\end{abstract}






\maketitle
\section{Introduction}
\label{sec:intro}

\begin{figure}
    \centering
    \includegraphics[width=0.9\linewidth]{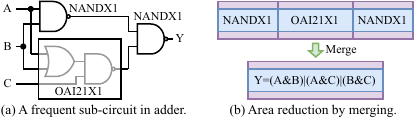}
    \vspace{-13pt}
    \caption{An example of reducing area by merging subcircuit.}
    \Description{A diagram showing an example of reducing area by merging frequent subcircuit.}
    \label{fig:intro}
\end{figure}

As integrated circuit manufacturing scales into deep submicron nodes, particularly at 22 nm and beyond, rising costs and physical constraints necessitate Design–Technology Co-Optimization (DTCO)~\cite{northrop2011design,liu20211,zhang20241}, where the standard cell library serves as the fundamental building block determining final circuit quality. 
The manual and time-consuming nature of cell layout design, however, becomes a bottleneck in DTCO iteration, driving the need for automated layout generation~\cite{guruswamy1997cellerity,togni2002automatic,majumder2011method,motiani2013method,karmazin2013celltk,ziesemer2014simultaneous,vaidyanathan2014sub,xu2017standard,cardoso2018libra,de2019transistor,van2019bonncell,jo2019design,lee2020tplace,li2020mcell,park2020sp,baek2021simultaneous,cheng2021probe2,ren2021nvcell}. 
To address process non-idealities and complex circuit demands, Standard Cell Library Extension (SCLX) has emerged as a key DTCO enabler. 
As shown in the simplified example in Figure~\ref{fig:intro}, by merging multiple basic cells into composite units, SCLX reduces gate-level redundancy and wiring congestion, thereby improving Quality of Results (QoR)~\cite{cheng2021complementary,li2022nctucell,choi2023probe3,liang2022autocelllibx,fu2025temacle}. 
Thus, developing an efficient and automated SCLX framework is essential for modern VLSI design automation.

Existing automated SCLX methods remain fundamentally constrained by their search space. Frameworks such as AutoCellLibX~\cite{liang2022autocelllibx} and TeMACLE~\cite{fu2025temacle} identify candidate cells via frequent subgraph mining on a post-mapping netlist, e.g., extracting $K$-feasible cones and verifying equivalence using SAT. 
However, these approaches operate only on a single post-mapping netlist from logic synthesis, restricting discovery to one topological realization of the Boolean function. 
This structural dependence subjects SCLX to a ``phase-ordering problem''~\cite{kulkarni2006exhaustive}: if a more efficient subcircuit is absent from the initial netlist, it remains inaccessible to subsequent mining.

To overcome SCLX's dependence on a fixed netlist, a paradigm shift toward systematic exploration of logical equivalence is needed. 
\textit{Equality saturation}, based on the \textit{e-graph} (equality graph) data structure, offers such a solution~\cite{tate2009equality}. 
The e-graph employs congruence closure to compactly encode equivalent expression structures through E-Classes~\cite{nieuwenhuis2005proof}. 
During saturation, rewrite rules expand the graph to represent all equivalent implementations, resolving the phase-ordering problem by deferring optimization until extraction~\cite{willsey2021egg}. 
This approach has demonstrated effectiveness in compiler optimization~\cite{zhao2024felix}, logic synthesis~\cite{chen2025morphic}, and symbolic reasoning~\cite{yin2025boole}. 
Integrating equality saturation into SCLX shifts the search from structural to semantic, enabling discovery of optimal cell patterns across logically equivalent netlists.

Based on the superior optimization capabilities of Equality Saturation, this paper proposes \papername{}: Automated Standard Cell Library Extension via Equality Saturation. As the first framework leveraging equality saturation for systematic standard cell enhancement, \papername{} introduces a semantic-driven methodology. Initially, equality saturation is applied to the post-mapping netlist, constructing an e-graph that compactly encodes all logically equivalent implementations. A modified subcircuit mining algorithm then operates on this e-graph, enabling discovery of high-frequency patterns across the full equivalence space. 
This approach reduces reliance on initial technology mapping decisions, thereby generating better cells to enhance overall circuit QoR.
The main \mbox{contributions of \papername{} are:}
\begin{itemize}[noitemsep, topsep=0pt]
    \item The first SCLX framework to resolve the phase-ordering problem via equality saturation, enabling semantic exploration across massive logically equivalent netlists.
    \item A novel subcircuit mining algorithm on e-graph that extracts optimal composite cells from compressed equivalence classes, unlocking patterns inaccessible to traditional netlist traversal.
    \item Empirical demonstration of significant area optimization, achieving a 15.41\% average area reduction and up to 23.64\% reduction compared to state-of-the-art tools.
    \item Confirmed effectiveness in a commercial flow, showing an average 8.00\% delay reduction and 1.27\% area reduction.
\end{itemize}

The remainder of this paper is organized as follows. Section~\ref{sec:pre} discusses the preliminaries, including the problem formulation of SCLX and the introduction of equality saturation. Section~\ref{sec:method} introduces the proposed SCLX framework. Section~\ref{sec:exp} details the experiment setting and evaluation of \papername{}, and concludes in Section~\ref{sec:conclusion}.

\section{Preliminaries}
\label{sec:pre}

This section is divided into two parts. 
First, we formalize the key concepts, including netlists and subcircuits, and present the Standard Cell Library Extension (SCLX) problem. 
Next, we introduce the relevant concepts and definitions pertaining to equality saturation.

\subsection{Problem Formulation}
\label{sec:pre_problem}

\subsubsection{Technology Mapping}

Technology mapping~\cite{testa2018logic,de2023logic,soeken2018epfl,brayton2010abc}, denoted by $\mathcal{M}$, is a graph covering problem transforming the technology independent Boolean network $B$ into an optimal, technology specific netlist $G$. The transformation is constrained by the Standard Cell Library $LIB$, a set of characterized physical resources. $\mathcal{M}$ is a complex optimization designed to minimize the Quality of Results (QoR) cost, $Q(G)$. This requires $G$ to be functionally equivalent to $B$ while containing only cells from $LIB$. Formally:

\begin{equation}
    \mathcal{M}_{LIB}(B) = \arg \min_{G} \{ Q(G) \mid G \sim LIB, \text{Func}(G) \equiv \text{Func}(B) \}.
\end{equation}

\subsubsection{Post-mapping Netlist}
The post-mapping netlist of a combinational circuit can be represented as a directed acyclic graph (DAG) $G(V, E, L)$ with a vertex set $V$, a directed edge set $E\subseteq V\times V$, a label mapping $L$, and no cycle. The definitions are shown below:
\begin{itemize}[noitemsep, topsep=0pt]
    \item Vertices $V$, also denoted as $V_G$, represent logic gates in the netlist, i.e., any vertex $v\in V$ is an instance of standard cell.
    \item Edges $E$, also denoted as $E_G$, represent nets from driven pins to sink pins in the netlist, i.e., any edge $e=(u,v)\in E$ represents gate $u$ drives gate $v$.
    \item Label mapping $L$, represents the labels of vertices and edges. Specifically, vertex label $L(v)$ means the standard cell type of $v$, while edge label $L(e)=L(u,v)=(O_u,I_v)$ for edge $e=(u,v)$ means the output pin named $O_u$ of gate $u$ drives the input pin named $I_v$ of gate $v$.
    \item Primary inputs $PI\in V$ are defined as vertices with no incoming edge, while primary outputs $PO\in V$ are defined as vertices with no outgoing edge. Specifically, $PI$ and $PO$ of a netlist are considered as specific vertices with labels \texttt{input} and \texttt{output}, with pins \texttt{O} and \texttt{I}, respectively.
\end{itemize}

\subsubsection{Subcircuits}
\label{sec:pre_subcircuit}
Subcircuit $S$ is a subgraph of netlist graph $G$. 
The input vertices and output vertices of $S$ are denoted by $SI$ and $SO$. 
The size of a subcircuit $|S|$ is defined as the number of vertices except primary inputs $|V_S-SI|$. 
The left most ellipse in Figure~\ref{fig:subcircuit_example} shows a simple example of a subcircuit. Here, $V_S=\{g_1, g_2, \ldots, g_6\}$, $SI=\{g_1 ,g_2, g_3\}$, $SO=\{g_5\}$. The main body of the subcircuit is $V_S-SI=\{g_4,g_5\}$ and hence $|S|=2$.
Moreover, a valid subcircuit must satisfy the following constraints: 
\begin{itemize}
    \item \textit{Single-output.} The same as TeMACLE~\cite{fu2025temacle}, in this work we only consider subcircuits with exactly one output, i.e., $|SO|=1$. 
    \item \textit{Connectivity.} All gates in $S$ should be connected.
    \item \textit{Source-sink.} Any input pin of gates $v\notin SI$ should be driven by one output pin of another gate $u$. Any output pin of gates $v\notin SO$ should drive at least one input pin of another gate $u$.
\end{itemize}
We define \textit{functional equivalent} of two subcircuits as follows:

\begin{figure}
    \centering
    \includegraphics[width=0.9\linewidth]{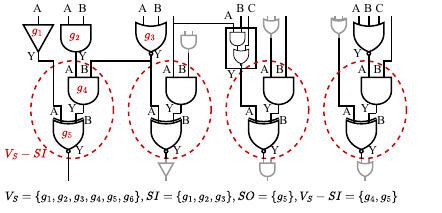}
    \vspace{-13pt}
    \caption{Example of a frequent subcircuit.}
    \Description{A diagram showing a subcircuit example.}
    \label{fig:subcircuit_example}
    \vspace{3pt}
\end{figure}

\begin{definition}[functional equivalent]
Consider two subcircuits $S_1,S_2$ with output vertices $SO_1=\{z_1\},SO_2=\{z_2\}$ and input vertices $SI_1=\{x_1,x_2,\ldots,x_m\},SI_2=\{y_1,y_2,\ldots,y_n\}$. Using boolean algebra, we can get boolean formulas for $z_1,z_2$, i.e., $z_1=f_1(SI_1)=f_1(x_1,x_2,\ldots,x_m)$, $z_2=f_2(SI_2)=f_2(y_1,y_2,\ldots,y_n)$. $S_1$ and $S_2$ is \textit{functional equivalent}, denoted by $S_1\equiv S_2$, if and only if $m=n$ and there exists a bijection $g: SI_1\xrightarrow{\sim}SI_2$ such that $f_1(SI_1)\equiv f_2(g(SI_1))$ for any boolean assignment $SI_1\in\{\mathbf{True}, \mathbf{False}\}^m$.
\end{definition}

\begin{figure*}
    \centering
    \includegraphics[width=0.95\linewidth]{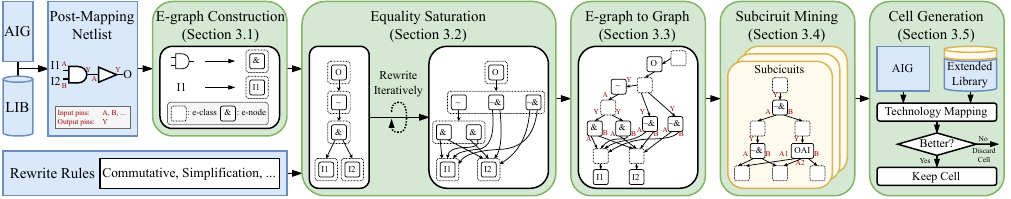}
    \vspace{-8pt}
    \caption{The workflow of proposed \papername{}.}
    \Description{A diagram showing the workflow of \papername{}.}
    \label{fig:workflow}
    \vspace{-7pt}
\end{figure*}

\subsubsection{Standard Cell Library Extension (SCLX)}
We further define patterns to conveniently describe the SCLX problem. A \textit{pattern group} $PG=\{S_1,S_2,\ldots,S_p\}$ is a set of $p$ functional equivalent subcircuits, i.e., $\forall S_i,S_j\in PG$, $S_i\equiv S_j$. Figure~\ref{fig:subcircuit_example} shows an example of pattern group with size 4.
Each pattern group $PG$ can be generated into a standard cell $c$ with tools like ASTRAN~\cite{ziesemer2014simultaneous}, AutoCellGen~\cite{baek2021simultaneous}. 
We denote the corresponding pattern group of $c$ as $PG_c$. Then, we have the following SCLX problem:

\noindent\textbf{Inputs:}
\begin{enumerate}[label=\arabic*), noitemsep, topsep=0pt]
    \item A given Boolean network $B$.
    \item An original standard cell library $LIB$.
    \item A technology mapping tool $\mathcal{M}$.
\end{enumerate}
\noindent\textbf{Outputs:} An extended standard cell library $LIB'=LIB\cup EX$, where $EX=\{c_1,c_2,\ldots,c_t\}$ is the set of extended cells.

\noindent\textbf{Constraints:}
\begin{enumerate}[label=\arabic*), noitemsep, topsep=0pt]
    \item At most $T$ new cells can be added to the extended library $LIB'$, i.e., $|EX|=t\leq T$.
    \item The number of original cells in any subcircuit is at most $N$, i.e., $\forall c_i\in EX$, $\forall S_j\in PG_{c_i}$, $|S_j|\leq N$.
    \item Any cell has at most $K$ input pins, i.e., $\forall c_i$, $\forall S_j\in PG_{c_i}$, $|SI|\leq K$.
    \item Any cell has one output pin, i.e., $\forall c_i$, $\forall S_j\in PG_{c_i}$, $|SO|=1$.
\end{enumerate}

\noindent\textbf{Goal:} Minimize the QoR cost function $Q(\mathcal{M}_{LIB'}(B))$ of the final mapped netlist $\mathcal{M}_{LIB'}(B)$: $\min_{LIB'}Q(\mathcal{M}_{LIB'}(B))$.

\subsection{Equality Saturation}
\label{sec:pre_egraph}

Equality saturation~\cite{nelson1980fast,nieuwenhuis2005proof,tate2009equality,willsey2021egg,yin2025boole} is a rewriting optimization technique powered by an e-graph (equivalence graph) data structure, which represents an equivalence relation over expressions.

\subsubsection{E-graph}
An \textit{e-graph} $\mathcal{G}$ is defined as $\mathcal{G}=(\mathcal{N},\mathcal{C},\mathcal{L})$, where:
\begin{itemize}[noitemsep, topsep=0pt]
    \item $\mathcal{N}$ is a set of vertices, called \textit{e-nodes}, where each e-node represents a distinct expression or sub-expression.
    \item $\mathcal{C}=\{C_1, C_2, \ldots, C_p\}$ is a partition of $\mathcal{N}$ into $p$ sets. Each e-node set $C_i$ represents an equivalence class known as an \textit{e-class}. E-nodes within the same e-class are considered equivalent.
    \item $\mathcal{L}:\mathcal{N}\to\Sigma\times\mathcal{C}^k$, called \textit{language}, is a labeling function that assigns each e-node to an operator and an ordered list of child e-classes, where $\Sigma$ represents the set of operators with $k$ operands and $\mathcal{C}^k$ denotes an ordered tuple of $k$ child e-classes from $\mathcal{C}$.
\end{itemize}

\begin{figure}
    \centering
    \includegraphics[width=0.9\linewidth]{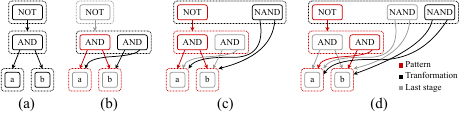}
    \vspace{-12pt}
    \caption{Simple e-graph rewriting example.}
    \Description{A diagram showing a simple e-graph example.}
    \label{fig:egraph_example}
\end{figure}

Figure~\ref{fig:egraph_example}(a) shows a simple e-graph example of $\text{NOT}(\text{AND}(a,b))$, 
where the arrows (from left to right) originating from each e-node represent the ordered child e-classes.

\subsubsection{Rewrite rules}
Rewrite rules $\mathcal{R}$ is a set of rules to modify the e-graph. Each rule is a pair of patterns $(l,r)$, also know as $l\to r$, which means the left pattern $l$ is equivalent to the right pattern $r$. Each pattern is a sub-expression like $\text{NOT}(x)$, $\text{AND}(x,y)$ and $\text{NAND}(x,y)$. To apply a rule, new e-nodes may be added and some e-classes may be extended or merged to represent new equivalence. Figure~\ref{fig:egraph_example}(b-d) shows the rewriting procedure of rules:
\begin{equation}
    \text{AND}(x,y)\to\text{AND}(y,x),\ 
    \text{NOT}(\text{AND}(x,y))\to\text{NAND}(x,y).
\end{equation}
In each step, the gray part remains unchanged, the red pattern is matched and the black transformation is applied.
\section{CellE}
\label{sec:method}

This section proposes an automated Standard Cell Library Extension (SCLX) framework via
equality saturation, namely \papername{}. 
The overall workflow is illustrated in Figure~\ref{fig:workflow}. 
The framework accepts a Boolean network as input, such as an And-Invertor Graph (AIG). Initially, \papername{} maps the Boolean network into a post-mapping netlist using given technology mapping tools. 
Subsequently, \papername{} converts the post-mapping netlist into an e-graph (Section~\ref{sec:method_construct}). 
Then, the rewrite engine of \papername{} applies rewriting rules to expand the e-graph until it is saturated (Section~\ref{sec:method_saturate}). 
Next, \papername{} converts the e-graph into a normal graph version (Section~\ref{sec:method_graph}), preparing for subciruit mining with graph mining algorithm (Section~\ref{sec:method_mine}). 
Finally, \papername{} generates new standard cells for mined subcircuits and minimizes the QoR cost function of the post-mapping netlist generated with the extended standard cell library (Section~\ref{sec:method_gen}). 


\begin{algorithm}[t]
\small
\caption{E-graph Construction}
\label{alg:egraph_construct}
\SetKwInOut{Input}{Input}
\SetKwInOut{Output}{Output}
\Input{Post-mapping netlist $G$}
\Output{E-graph $\mathcal{G}$}
Initialize empty e-class mapping $vmap$ and e-graph $\mathcal{G}$\;
$order\gets \text{TopoSort}(G)$\tcp*{\small From inputs to outputs}
\For{each vertex $v$ in $order$}{
    $inputs \gets [\,]$\tcp*{\small Collect input e-classes}
    \For{each input $i$ of vertex $v$}{
        $inputs.\text{push}(vmap[i])$\tcp*{\small Same pin order}
    }
    $cid \gets \mathcal{G}.\text{insert}(v, inputs)$\tcp*{\small Add e-node}
    $vmap[v] \gets cid$\;
}
\end{algorithm}


\subsection{E-graph Construction}
\label{sec:method_construct}

After mapping the Boolean network into a post-mapping netlist $G$, \papername{} constructs an e-graph $\mathcal{G}$ in preparation for subsequent equality saturation. 
The algorithm for egraph construction is detailed in Algorithm~\ref{alg:egraph_construct}.
\papername{} first computes a topological order of the netlist vertices, from primary inputs $PI$ to primary outputs $PO$. It then iterates through the vertices in this order. 
For each vertex $v$, it gathers the e-class IDs of its inputs using a mapping $vmap$. The input order for each e-node corresponds to the pin order for each gate.
Finally, it adds a new e-node representing $v$ and its inputs to the e-graph $\mathcal{G}$, and updates $vmap$ with the resulting e-class ID. 
The complexity is $O(|V_G|+|E_G|)$.

\begin{table}
\centering
\caption{Example of Rewrite Rules.}
\vspace{-8pt}
\label{tab:rules}
\resizebox{0.9\linewidth}{!}{%
\begin{tabular}{c c c}
\toprule
\textbf{Name} & \textbf{Pattern (LHS)} & \textbf{Transformation (RHS)} \\
\midrule
\multirow{2}{*}{Commutativity} & $\text{AND}(a, b)$ & $\text{AND}(b, a)$ \\
                               & $\text{XOR}(a, b)$ & $\text{XOR}(b, a)$ \\
\midrule
\multirow{2}{*}{De Morgan's} & $\text{OR}(\text{NOT}(a), \text{NOT}(b))$ & $\text{NAND}(a, b)$  \\
                             & $\text{AND}(\text{NOT}(a), \text{NOT}(b))$ & $\text{NOR}(a, b)$  \\
\midrule
\multirow{2}{*}{Simplification} & $\text{NOT}(\text{AND}(a, b))$ & $\text{NAND}(a, b)$ \\
                                & $\text{NOT}(\text{XOR}(a, b))$ & $\text{XNOR}(a, b)$ \\
\midrule
Involution & $\text{NOT}(\text{NOT}(a))$ & $a$ \\
\bottomrule
\end{tabular}
} %
\vspace{1pt}
\end{table}

\begin{algorithm}[t]
\small
\caption{E-graph to Normal Graph}
\label{alg:egraph_to_graph}
\SetKwInOut{Input}{Input}
\SetKwInOut{Output}{Output}
\Input{E-graph $\mathcal{G}=(\mathcal{N},\mathcal{C},\mathcal{L})$, library pins $(\mathcal{I},\mathcal{O})$}
\Output{Graph $G=(N,C,E)$}
Initialize empty graph $G$\;
Create vertex mapping $M_\mathcal{N},M_\mathcal{C}$ for $\mathcal{N}$ and $\mathcal{C}$, respectively\;
\For{e-class $C\in\mathcal{C}$}{
    \For{e-node $n\in C$ with output pin $\mathcal{O}(n)$}{
        Add edge: $M_\mathcal{C}[C] \to M_\mathcal{N}[n]$ with label $\mathcal{O}(n)$\;
    }
}
\For{e-node $n\in\mathcal{N}$}{
    \For{e-class and pin $(C,I)\in \text{zip}(\mathcal{L}(n),\mathcal{I}(n))$}{
        Add edge: $M_\mathcal{N}[n] \to M_\mathcal{C}[C]$ with label $I$\;
    }
}
\end{algorithm}

\subsection{Equality Saturation}
\label{sec:method_saturate}

The rewriting procedure utilizes equality saturation to exhaustively explore equivalent circuit representations. This iterative process repeatedly applies a predefined set of rewrite rules, such as the examples shown in Table~\ref{tab:rules}, to the e-graph $\mathcal{G}$. When a rule's left-hand side (pattern) is matched within the e-graph, the corresponding right-hand side (transformation) is added to the same e-class, effectively asserting their equivalence. \papername{} continues this application until saturation is reached, meaning no new equivalent structures can be added. The resulting saturated e-graph $\mathcal{G}'$ thus compactly encodes a vast space of functionally equivalent circuits. The complexity is $O(r|\mathcal{N}|^k)$ with $r$ rules and maximal pattern size $k$.

\subsection{E-graph to Graph Conversion}
\label{sec:method_graph}

\begin{figure}
    \centering
    \includegraphics[width=\linewidth]{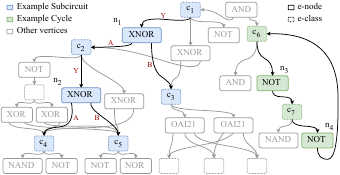}
    \vspace{-18pt}
    \caption{Example of graph version e-graph.}
    \Description{A diagram showing an example of graph version e-graph.}
    \label{fig:egraph_subcircuit}
    \vspace{2pt}
\end{figure}

\papername{} notes that the e-graph $\mathcal{G}$ is an unusual data structure, primarily designed for equivalence management rather than direct graph mining. Specifically, it lacks explicit gate pin information and represents alternatives compactly, making it unsuitable for direct application of subcircuit mining algorithms.
Therefore, \papername{} introduces Algorithm~\ref{alg:egraph_to_graph} to transform the saturated E-graph $\mathcal{G}=(\mathcal{N},\mathcal{C},\mathcal{L})$ into a standard graph $G$ suitable for mining. The algorithm initializes $G$ and creates corresponding vertices for all e-nodes $\mathcal{N}$ and e-classes $\mathcal{C}$, recording these mappings as $M_\mathcal{N}$ and $M_\mathcal{C}$, respectively. The conversion proceeds in two main stages, incorporating the library pin information (inputs $\mathcal{I}$ and outputs $\mathcal{O}$). First, for every e-class $C$, directed edges are added from the vertex representing $C$ (via $M_\mathcal{C}$) to all contained e-nodes $n$ (via $M_\mathcal{N}$), labeled with the corresponding output pin $\mathcal{O}(n)$. Second, for every e-node $n$, directed edges are added from the vertex representing $n$ via ($M_\mathcal{N}$) to the vertices representing its child e-classes $C$ (via $M_\mathcal{C}$), labeled with the respective input pin $I$ (obtained from the language $\mathcal{L}(n)$ and $\mathcal{I}(n)$). This yields a bipartite graph structure $G$ with explicit connectivity and pin labeling. Figure~\ref{fig:egraph_subcircuit} shows an example of converted graph. The complexity is $O(|V_G|+|E_G|)$.

\begin{algorithm}[t]
\small
\caption{Pattern Groups Collection}
\label{alg:mining}
\SetKwInOut{Input}{Input}
\SetKwInOut{Output}{Output}
\SetKwFunction{SubgraphMining}{SubgraphMining}
\SetKwFunction{FOne}{FindFrequent1EdgeWithProjections}
\SetKwFunction{IsMin}{IsMinimal}
\SetKwFunction{IsIllegal}{IsIllegal}
\SetKwFunction{Constr}{SatisfiesConstraints}
\SetKwFunction{Exts}{FindExtensions}
\Input{Graph $G$, MinSupport $\theta_{min}$}
\Output{Frequent Patterns $\mathcal{F}$}
$\mathcal{F} \gets \emptyset$\;
$E_{freq} \gets$ \FOne{$G, \theta_{min}$}\label{alg:mining_one}\;
\For{$(code, P) \in E_{freq}$}{
    \SubgraphMining{$G$, $\theta_{min}$, $\mathcal{F}$, $code$, $P$}\;
}
\Return $\mathcal{F}$\;
\BlankLine
\SetKwProg{Fn}{Function}{}{end}
\Fn{\SubgraphMining{$G$, $\theta_{min}$, $\mathcal{F}$, $code$, $P$}}{
    \If{(not \IsMin{$code$}) or \IsIllegal{$code$}\label{alg:mining_min_start}}{
        \Return\tcp*{\small Prune non-canonical or illegal paths}\label{alg:mining_min_end}
    }
    \If{\Constr{$code$}\label{alg:mining_constr_start}}{
         $\mathcal{F} \gets \mathcal{F} \cup \{ (code, P) \}$\;\label{alg:mining_constr_end}
    }
    $Exts \gets$ \Exts{$G, \theta_{min}, code, P$}\;\label{alg:mining_ext_start}
    \For{$(code', P') \in Exts$}{
         \SubgraphMining{$G$, $\theta_{min}$, $\mathcal{F}$, $code'$, $P'$}\;\label{alg:mining_ext_end}
    }
}
\end{algorithm}

\subsection{Subcircuit Mining}
\label{sec:method_mine}

Following the conversion from the e-graph, \papername{} obtains a normal, bipartite graph $G$ that explicitly represents both gates (e-nodes) and equivalence points (e-classes) with explicit pin labeling. 
To identify frequently occurring and valuable subcircuits (or pattern groups), \papername{} employs a subcircuit mining phase. 
This phase uses a frequent subgraph mining algorithm, adapted from the gSpan~\cite{yan2002gspan} method, which is renowned for discovering canonical subgraph patterns in general graphs (even with cycles).
Note that the saturated e-graph $\mathcal{G}$ may contain cycles and thus the converted graph $G$ (e.g., green parts in Figure~\ref{fig:egraph_subcircuit}) also may. Therefore, mining methods designed for DAGs, like $K$-feasible cones in TeMACLE~\cite{fu2025temacle} are not suitable.

\begin{table*}[htbp]
\centering
\caption{Experimental results on the EPFL benchmark~\cite{amaru2015epfl} and FreePDK45 library~\cite{stine2007freepdk}.}
\vspace{-10pt}
\resizebox{0.93\linewidth}{!}{%
\begin{tabular}{c|rrr|rrrr|rrrrrr}
\toprule
\multirow{2}{*}{Circuit} & \multicolumn{3}{c|}{Original} & \multicolumn{4}{c|}{TeMACLE~\cite{fu2025temacle}} & \multicolumn{6}{c}{\papername{}} \\
\cmidrule{2-14}
 & gates & area ($\mu m^2$) & depth & area ($\mu m^2$) & reduction (\%) & depth & size & e-nodes & area ($\mu m^2$) & reduction (\%) & depth & time (s) & size \\
\midrule
adder128 & 768 & 2373.25 & 193 & 1884.15 & 20.61 & 129 & 3/5 & 2037 & 1438.80 & 39.37 & 128 & 6.04 & 2/3 \\
bar & 2283 & 5426.52 & 10 & 4607.51 & 15.09 & 9 & 3/2/2/4 & 6521 & 4483.56 & 17.38 & 9 & 18.90 & 2/3/3/2/3 \\
cavlc & 532 & 1232.38 & 11 & 1187.73 & 3.62 & 10 & 2/2/2/2/2 & 1721 & 1186.28 & 3.74 & 11 & 0.21 & 2/2/2/4/4 \\
div & 42220 & 109675.88 & 2213 & 86576.50 & 21.06 & 2200 & 2/2/2/2/3 & 100550 & 84004.44 & 23.41 & 2190 & 444.13 & 2/2/3 \\
hyp & 187027 & 505059.72 & 10749 & 415107.84 & 17.81 & 10733 & 2/2/2/2/2 & 485906 & 413745.44 & 18.08 & 10695 & 136.75 & 2/2/2/4/2 \\
i2c & 1094 & 2461.95 & 11 & 2368.51 & 3.80 & 11 & 2/2/3/2/3 & 3796 & 2327.08 & 5.48 & 11 & 0.77 & 2/2/4/2/2 \\
int2float & 182 & 420.49 & 9 & 407.74 & 3.03 & 9 & 2/2/3 & 635 & 402.35 & 4.31 & 9 & 0.37 & 2/2/2/2/2 \\
log2 & 21827 & 58706.61 & 232 & 52795.23 & 10.07 & 221 & 2/2/2/2/2 & 65421 & 52868.23 & 9.95 & 221 & 27.86 & 2/2/2/2/2 \\
max & 2694 & 6006.57 & 178 & 5002.03 & 16.72 & 176 & 2/2/3/2/2 & 6092 & 4901.63 & 18.40 & 177 & 12.77 & 2/2/2/3/3 \\
mem\_ctrl & 36517 & 81936.96 & 74 & 73855.93 & 9.86 & 74 & 2/2/2/3/2 & 113405 & 79494.13 & 2.98 & 74 & 31.25 & 2/2/2/2/2 \\
multiplier & 16684 & 48909.51 & 179 & 43687.63 & 10.68 & 178 & 2/2/2/3/3 & 38048 & 43483.38 & 11.09 & 178 & 10.54 & 2/2/2/2/2 \\
priority & 1072 & 2344.15 & 64 & 1691.72 & 27.83 & 64 & 2/2/3/2 & 2751 & 2133.25 & 9.00 & 64 & 20.37 & 2/2/2 \\
router & 240 & 569.73 & 25 & 542.09 & 4.85 & 25 & 2/2/3/2/2 & 821 & 540.23 & 5.18 & 25 & 2.94 & 2/2/2/2 \\
sin & 4145 & 10882.60 & 116 & 9609.23 & 11.70 & 113 & 2/2/2/2/3 & 12603 & 9285.37 & 14.68 & 113 & 4.97 & 2/2/2/2/2 \\
sqrt & 18140 & 47752.68 & 3039 & 32940.58 & 31.02 & 3038 & 2/2/2/5/4 & 55225 & 33461.75 & 29.93 & 3039 & 944.94 & 2/2/2/2 \\
square & 13895 & 37336.10 & 167 & 33215.22 & 11.04 & 167 & 2/2/2/4/2 & 35789 & 30100.20 & 19.38 & 125 & 1.68 & 2/2/3/2/2 \\
voter & 9973 & 28922.49 & 38 & 23512.72 & 18.70 & 33 & 2/2/2/3/2 & 25064 & 23311.05 & 19.40 & 32 & 2.75 & 2/2/3/4 \\
\midrule
gmean & - & 11425.74 & 114.36 & 9785.17 & 14.36 & 108.82 & - & - & 9665.16 & 15.41 & 107.33 & - & - \\
\bottomrule
\end{tabular}
} %
\label{tab:gscl_results}
\vspace{-10pt}
\end{table*}

\subsubsection{Subgraph Constraints}
\label{sec:method_constraints}
It is important to note that in this bipartite graph $G(N,C,E,L)$---where $N$ is the e-node set, $C$ is the e-class set, $E$ is the directed edge set, and $L$ is the label mapping---the definition of a subcircuit is different from that in Section~\ref{sec:pre_subcircuit}. 
A subcircuit is a subgraph $S$ that satisfies the following constraints:
\begin{itemize}[noitemsep, topsep=0pt]
    \item \textbf{\textit{Completeness.}} All input $SI$ and output $SO$ vertices must be e-class vertices, to ensure that all pins of each gate are included in the subgraph, i.e., $SI,SO\subseteq C_S$.
    \item \textbf{\textit{Uniqueness.}} Only one e-node can be chosen in any e-class, i.e., any e-class $c\in C_S-SO$ must have one outgoing edge.
    \item \textit{Single-output.} The same as TeMACLE~\cite{fu2025temacle}, $S$ has exactly one output vertex, i.e., $|SO|=1$.
    \item \textit{Connectivity.} All vertices in $S$ should be connected.
    \item \textit{Source-sink.} All pins of a gate must either drive another gate or be driven by another gate, i.e., any e-node $n\in N_S$ has exactly one incoming edge and all its outgoing edges are within $E_S$.
\end{itemize}

The blue part in Figure~\ref{fig:egraph_subcircuit} shows an example of a subcircuit $S$ with e-class inputs $SI=\{c_3,c_4,c_5\}$ and outputs $SO=\{c_1\}$. 

\subsubsection{Subcircuit Encoding}
\label{sec:method_encoding}

To efficiently manage and compare subgraphs, \papername{} uses a canonical representation based on a Depth-First Search (DFS) code, as inspired by gSpan~\cite{yan2002gspan}. A subgraph pattern is encoded as a \texttt{DFSCode}, which is defined as a sequence of \texttt{DFSEdge}s. Each \texttt{DFSEdge} is a 5-tuple $(i, j, L_i, L_{ij}, L_j)$, representing an edge from pattern vertex $i$ to pattern vertex $j$, with node labels $L_i$ and $L_j$ (either an e-class or an e-node with its cell type) and an edge label $L_{ij}$ (the pin name).

The sequence of edges in the \texttt{DFSCode} corresponds to a specific DFS traversal of the subgraph. 
By using a \textit{DFS lexicographic order} on these sequences of 5-tuples, \papername{} can generate a unique, minimal \texttt{DFSCode} for any given subgraph topology. This canonical encoding is critical, as it allows for direct comparison of patterns and pruning of redundant search paths.
A \textit{pattern group} is thus defined not just by its canonical \texttt{DFSCode} (the pattern's structure), but also by its associated set of \textit{projections}---a list of all mappings from the pattern's nodes to the nodes in the main graph $G$ where that specific pattern occurs. Each projection corresponds to a subcircuit. The size of this projection set is the pattern's \textit{support}.

\subsubsection{Pattern Groups Collection}
\label{sec:method_collection}

The collection of frequent pattern groups is performed using a depth-first search on the lattice of possible subgraphs, as detailed in Algorithm~\ref{alg:mining}. The process begins by identifying all single-edge patterns (1-edge \texttt{DFSCode}s) in the graph $G$ that meet a minimum support threshold, $\theta_{min}$ (line \ref{alg:mining_one}). These initial patterns, along with their projection lists, form the starting point for the recursive mining process.

The core of the algorithm is the recursive \texttt{SubgraphMining} function. This function takes a \texttt{DFSCode} and its list of projections $P$. 
First, it checks if the \texttt{DFSCode} is in its minimal (line \ref{alg:mining_min_start}-\ref{alg:mining_min_end}) canonical form using DFS lexicographical ordering to avoid redundant exploration of the same subgraph discovered via different traversal paths. 
Next, \papername{} checks if the pattern satisfies structural constraints, such as the maximum size, the number of inputs and constraints from Section~\ref{sec:method_constraints} (line \ref{alg:mining_min_start}-\ref{alg:mining_constr_end}). 
Valid and frequent patterns are saved to set $\mathcal{F}$. 
The function then recursively attempts to extend the current pattern by one edge (line \ref{alg:mining_ext_start}-\ref{alg:mining_ext_end}). 
This process continues until no further frequent extensions can be found. The complexity is $O(|E_G|\times|V_G|^N)$ where $N$ is the maximal subcircuit size.

\subsection{Standard Cell Generation}
\label{sec:method_gen}

The final stage involves generating and selecting new standard cells from the frequent pattern groups $\mathcal{F}$ identified in Section~\ref{sec:method_mine}. Each pattern group is treated as a candidate standard cell. Following the approach of TeMACLE~\cite{fu2025temacle}, \papername{} first calculates each cell's minimized function expression using the Quine-McCluskey algorithm~\cite{quine1955way,mccluskey1956minimization}. Subsequently, \papername{} uses SAT to filter functional equivalent subcircuits. Afterward, \papername{} generates SPICE netlists and employs automated tools, such as ASTRAN~\cite{ziesemer2014simultaneous} and AutoCellGen~\cite{baek2021simultaneous}, to perform standard cell generation. 
Since it is impractical to generate and include cells for all mined patterns into the library, a selection phase is required. \papername{} evaluates the utility of the candidate cells by performing technology mapping for the original Boolean network using the original library augmented with the new candidates. The primary objective is to identify the subset of candidate cells that minimizes a predefined QoR cost function. This selected set of high-value cells forms the final extended standard cell library, which is the ultimate output of the \papername{} framework.
\section{Experimental Results}
\label{sec:exp}


The experiments were conducted on a system with a Dual-Socket AMD EPYC 9754 128-Core Processor @ 1.50GHz and 1.5 TiB of memory. 
\papername{} is implemented using Rust language for cell finding (the translation between graphs and e-graphs, equality saturation and subcircuit mining), and using Python language for the cell generation. 
The \textit{egg}~\cite{willsey2021egg} framework was used for equality saturation. For fair area comparison, the experiment chose the same setup as TeMACLE~\cite{fu2025temacle}, which were the EPFL benchmark~\cite{amaru2015epfl}, the synthesis tool mockturtle~\cite{soeken2018epfl}, the standard cell layout generation tool ASTRAN~\cite{ziesemer2014simultaneous} and the original standard cell library FreePDK45~\cite{stine2007freepdk} with standard cells AND2X2, AOI21X1, BUFX2, INVX1, NAND2X1, NAND3X1, NOR2X1, NOR3X1, OAI21X1, OR2X2, XNOR2X1, and XOR2X1. Following the setup of TeMACLE, the maximal number of extended standard cells $T$ was set to 5. The maximal number of original cells in any new cell $N$ was set to 5. The maximal number of inputs for any new cell $K$ was set to 3. For rewrite rules, \papername{} chose 14 commulative, 4 De Morgan's, 6 simplication, and 1 involution.

\begin{figure}
    \centering
    \includegraphics[width=0.9\linewidth]{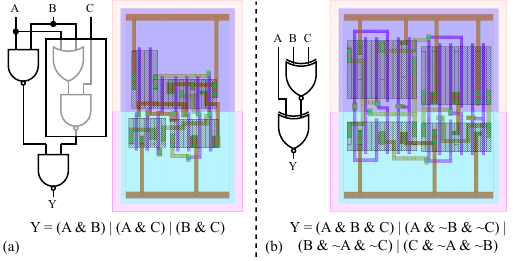}
    \vspace{-10pt}
    \caption{New standard cells generated for ``adder128''.}
    \Description{A diagram showing new standard cells generated for ``adder128''.}
    \label{fig:adder}
    \vspace{3pt}
\end{figure}

\subsection{Area Comparison}
\label{sec:exp_area}

Table \ref{tab:gscl_results} details the area results, comparing the Original baseline library mapping against the extended libraries from TeMACLE and \papername{}. The table lists the final area, percentage area reduction (reduction (\%)), depth, and new cell composition with cell sizes (size). For \papername{}, the number of e-nodes (e-nodes) and the total runtime (time (s)) of its subcircuit finding stages are also listed.

On average, \papername{} achieves a 15.41\% geometric mean (gmean) area reduction over the original library, surpassing TeMACLE's 14.36\%. This highlights the effectiveness of our equality-saturation-based approach. The benefit is pronounced on specific circuits; for instance, on ``adder128'', \papername{} achieves a 23.64\% further area reduction compared to TeMACLE's result. For a few other circuits, the improvements are smaller because their logic exposes fewer semantic alternatives, while TeMACLE’s additional iterative remapping can occasionally surface patterns that align particularly well with its mining procedure. The number of e-nodes remains roughly three times the number of gates, keeping the e-graph size bounded. Moreover, subcircuit finding runtime is negligible or comparable to cell generation time (tens to hundreds of seconds per cell).

This ``adder128'' case demonstrates our advantage. TeMACLE identified 3-input MAJ and XOR subcircuits that included extra, area-increasing inverters, likely artifacts favored by the initial mapping. In contrast, \papername{}'s equality saturation engine explored equivalent representations and identified the more compact, \textit{non-inverted} 3-input MAJ and XOR functions as optimal candidates (function and layouts are shown in Figure~\ref{fig:adder}). While the inverted versions might seem preferable when built from discrete gates, the non-inverted versions discovered by \papername{} are significantly more area-efficient when generated as single, compact standard cells. This demonstrates that exploring functional equivalences is crucial for identifying the most beneficial cells.

\subsection{Applicability}
\label{sec:exp_compatibility}

\begin{figure}
    \centering
    \includegraphics[width=0.9\linewidth]{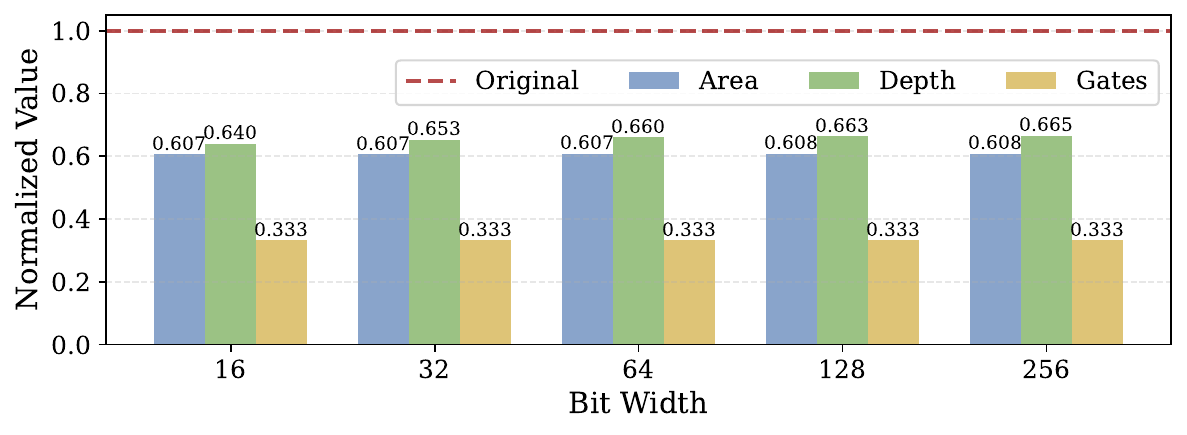}
    \vspace{-10pt}
    \caption{QoR reduction for adders with varying bit widths using original and \papername{} extended standard cell library.}
    \Description{A diagram showing QoR reduction of adders.}
    \label{fig:adder_qor}
\end{figure}

We tested the extended library on adders of varying scales to evaluate the general applicability of the cells mined from ``adder128''. As shown in Figure~\ref{fig:adder_qor}, the normalized area, depth, and gate count reductions are highly consistent for bit widths from 16 to 256. The metrics remain stable, with area at \textasciitilde 0.608, depth at \textasciitilde 0.66, and gates at 0.333. This demonstrates that the discovered cells possess strong applicability,
providing stable benefits across different circuit sizes.

\begin{figure}
    \centering
    \includegraphics[width=0.9\linewidth]{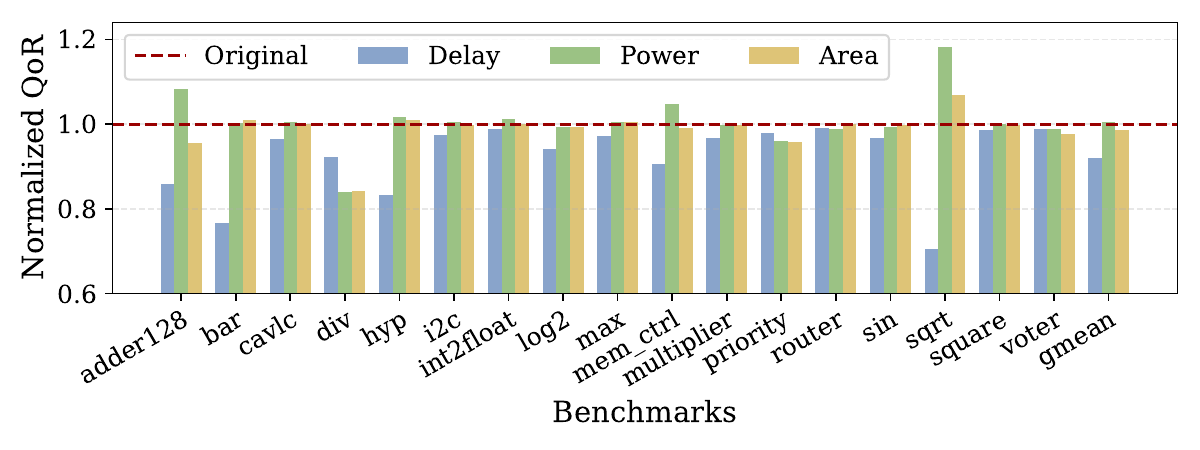}
    \vspace{-10pt}
    \caption{QoR reduction for EPFL benchmark~\cite{amaru2015epfl} using original ASAP7~\cite{vashishtha2017asap7} and \papername{} extended standard cell library.}
    \Description{A diagram showing QoR reduction of asap7.}
    \label{fig:asap7}
    \vspace{3pt}
\end{figure}

\subsection{Complete Characterization}
\label{sec:exp_char}

To further evaluate our approach, we used \papername{} with cell generation tool AutoCellGen~\cite{baek2021simultaneous} to extend ASAP7 7.5-track library~\cite{vashishtha2017asap7} (using the same cells as FreePDK45~\cite{stine2007freepdk}) and characterized the generated cells with Cadence Liberate 19.2.1.215. Then Cadence Genus 19.12-s121\_1 was used for logical synthesis. Figure~\ref{fig:asap7} shows the delay, power and area for EPFL benchmark~\cite{amaru2015epfl} using original library and the best extended library found by \papername{}. The QoR cost function was chosen as $\text{Delay}^2\times \text{Power}\times \text{Area}$, since delay is the most critical objective in general VLSI flow. As shown in Figure~\ref{fig:asap7}, the extended library achieves delay reduction on all benchmarks and also reduces area. On average (gmean), delay decreased by 8.00\% and area decreased by 1.27\%, while power remained unchanged. This is because cell fusion simplified logic and reduced depth, lowering delay. Furthermore, the commercial synthesis tool, usually optimizing delay first, prioritizes this delay reduction. This demonstrates \papername{}'s effectiveness in a standard commercial VLSI flow.
\section{Conclusion}
\label{sec:conclusion}

This paper presented \papername{}, a novel standard cell library extension framework utilizing equality saturation to exhaustively explore functionally equivalent subcircuits. By converting a post-mapping netlist into an e-graph, \papername{} enables the superior discovery of highly optimized cell candidates. We introduced a customized mining algorithm to efficiently select the most beneficial patterns. Experimental results on EPFL benchmarks demonstrate \papername{}'s effectiveness, achieving a 15.41\% average area reduction over the original library, and showcasing an area reduction of up to 23.64\% over prior work. Furthermore, when characterized in a commercial flow using ASAP7, the extended library shows an 8.00\% average reduction in delay, confirming \papername{}'s practical utility for QoR optimization.
Future work will focus on integrating physical cell properties into the equality saturation and mining to further optimize for QoR.

\clearpage
\bibliographystyle{ACM-Reference-Format}
\bibliography{references}










\end{document}